\documentclass[journal]{IEEEtran}
\IEEEoverridecommandlockouts
\usepackage{amsmath}
\usepackage{graphicx}
\usepackage{amssymb}
\usepackage{cases}
\usepackage{cite}
\usepackage{algpseudocode}
\usepackage{algorithm}
\usepackage{relsize}
\usepackage[nodisplayskipstretch]{setspace}

\begin{document}

\title{The Cost of Mitigating Power Law Delay in Random Access Networks}

\author{Suzhi~Bi,~\IEEEmembership{Student Member,~IEEE} and Ying~Jun~(Angela)~Zhang,~\IEEEmembership{Senior Member,~IEEE}\\
Department of Information Engineering, The Chinese University of Hong
        Kong,\\Shatin, New Territories, Hong Kong. \\ Email: \{bsz009, yjzhang\}@ie.cuhk.edu.hk
        \thanks{This work was supported in part by the Competitive Earmarked Research Grant (Project Number $419509$) established under the University Grant Committee of Hong Kong and Direct Research Grant (Project Number $2050439$) established under The Chinese University of Hong Kong.}}
\maketitle
\vspace{-0.7in}
\begin{abstract}
Exponential backoff (EB) is a widely adopted collision resolution mechanism in many popular random-access networks including Ethernet and wireless LAN (WLAN). The prominence of EB is primarily attributed to its asymptotic throughput stability, which ensures a non-zero throughput even when the number of users in the network goes to infinity. Recent studies, however, show that EB is fundamentally unsuitable for applications that are sensitive to large delay and delay jitters, as it induces divergent second- and higher-order moments of medium access delay. Essentially, the medium access delay follows a power law distribution, a subclass of heavy-tailed distribution. To understand and alleviate the issue, this paper systematically analyzes the tail delay distribution of general backoff functions, with EB being a special case. In particular, we establish a tradeoff between the tail decaying rate of medium access delay distribution and the stability of throughput. To be more specific, convergent delay moments are attainable only when the backoff functions $g(k)$ grows slower than exponential functions, i.e., when $g(k)\in o\left(r^k\right)$ for all $r>1$. On the other hand, non-zero asymptotic throughput is attainable only when backoff functions grow at least as fast as an exponential function, i.e., $g(k)\in\Omega\left(r^k\right)$ for some $r>1$. This implies that bounded delay moments and stable throughput cannot be achieved at the same time. For practical implementation, we show that polynomial backoff (PB), where $g(k)$ is a polynomial that grows slower than exponential functions, obtains finite delay moments and good throughput performance at the same time within a practical range of user population. This makes PB a better alternative than EB for multimedia applications with stringent delay requirements.
\end{abstract}


\section{INTRODUCTION}
Binary exponential backoff (BEB) is widely adopted as a key collision resolution mechanism in popular random-access networks, such as IEEE $802.3$ Ethernet and IEEE $802.11$ wireless local area network (WLAN). With exponential backoff (EB), a packet is transmitted after waiting a number of time slots randomly selected from a contention window, the size of which increases multiplicatively on collisions. Mathematically, the contention window $W_k=g(k)W_0$ after $k$ consecutive collisions of a packet. Here, $g(k)=r^k$ with $r>1$ is the backoff function for EB \footnote{Note that $g(k)$ must be an increasing function for the backoff process to be meaningful. Therefore, $r$ must be larger than unity.} and $W_0$ is the initial contention window size. BEB is a special case with $r=2$.

Most of the research attention has been focused on investigating the throughput provided by EB. Thanks to the seminal work of Bianchi\cite{2000:Bianchi}, the throughput is now well understood through a fixed point equation that characterizes the backoff process. Subsequently, \cite{2005:Kwok} shows that the throughput of EB is stable against the network size in the sense that the throughput converges to a nonzero constant when the network size goes to infinity (assuming no retry limit is enforced). Throughput stability has been the most intriguing aspect of EB, and has enabled EB-based MAC protocols to support a wide range of throughput oriented applications regardless of the network congestion level.

\begin{figure}
\centering
  \begin{center}
    \includegraphics[width=3.5in]{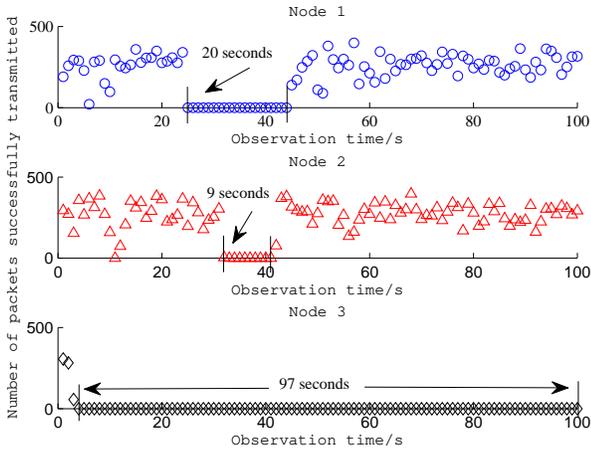}
  \end{center}
  \caption{Illustration of transmission starvation: number of packets transmitted in $100$ consecutive seconds for $3$ of $10$ nodes in a 802.11g system. Assume that all nodes are continuously backlogged and no retry limit is enforced.}
  \label{51}
\end{figure}

With the recent boom of delay-sensitive multimedia applications such as VoIP and video conferencing, research interests are being shifted to other aspects of system performance such as delay, delay jitter, and short-term fairness \cite{2009:Cho}. Indeed, it can be shown that delay jitter significantly affects the users' perception of quality of real-time multimedia services. EB, despite its good throughput performance, has been shown to suffer poor performance in delay and short-term fairness. More specifically, EB could induce divergent (i.e., infinite) second- and high-order moments of medium access delay, yielding extraordinarily large delay jitter and severe transmission starvation of users \cite{2007:Sakurai,2009:Liew,2009:Cho,2010:Zhang}. Essentially, the medium access delay follows a power law distribution, implying that a non-negligible number of packets may experience much larger delay than the average \cite{2009:Liew,2007:Sakurai,2009:Cho}. As a motivating example, we monitor the packet transmission during a $100$ second period in a $10$-node IEEE $802.11$g WLAN, where BEB is adopted. Alarmingly, $3$ out of the $10$ nodes experience severe transmission starvation, as illustrated in Fig. $\ref{51}$. The figure shows that node $1$ and $2$ perceive starvation for a duration of $20$ and $9$ seconds, respectively. Even worse, node $3$ barely receives any service throughout the entire simulation time.

In an attempt to address the above issues, this paper seeks to understand the following important questions.
\begin{enumerate}
  \item[\textbf{Q1}:] What is the root cause of the power-law delay distribution of EB. Is it an intrinsic issue of EB, or can be avoided by adjusting the backoff exponent $r$.
  \item[\textbf{Q2}:] If the problem is intrinsic with EB, can we find an alternative backoff function that does not suffer the same problem. In general, what is the necessary and sufficient condition for a backoff function to have convergent delay moments, i.e., not to experience power law delay.
  \item[\textbf{Q3}:] Is it possible to achieve throughput stability and convergent delay moments at the same time by certain backoff functions. If not, are there any backoff functions that exhibit convergent delay moments and good throughput performance at the same time when the network size is within a finite and practical range.
\end{enumerate}

In the literature, $\textbf{Q1}$ has been partly addressed. \cite{2007:Sakurai} first finds that the medium access delay distribution of EB is heavy-tailed when retry limit $K$ is infinite, regardless of the backoff exponent $r$. \cite{2009:Cho} later proves that the medium access delay indeed follows a power law distribution, the slope of which is obtained as a function of the backoff exponent and the collision probability. Noticeably, the effect of power law delay cannot be eliminated even if a finite retry limit K is enforced in practical systems. \cite{2009:Cho} and \cite{2007:Sakurai} observe that the medium access delay follows a truncated power law distribution, implying that small retry limit does not eliminate the power law characteristics induced by EB. This directly translates to high packet loss rate, if packets have to be discarded upon reaching a retry limit $K$. Indeed, our simulation results show that BEB suffers $10\%$ packet loss rate in a $50$-node network with $K=5$, leading to an equal percentage reduction of throughput as that with $K=\infty$. The analysis in these prior work can be treated as a special case of the analysis for general backoff functions in this paper.

As to $\textbf{Q2}$, there are some initial attempts to replace EB with other more moderate backoff algorithms, such as linear backoff ($g(k)=1+k$) and polynomial backoff ($g(k)=1+k^b,\ b>0$)\cite{1987:Hastad,2008:Xu,2011:Sun}. Observations made by \cite{2008:Xu} showed that linear and polynomial backoffs with appropriate parameter settings can improve upon BEB in terms of throughput and delay performance. \cite{2011:Sun} observed that PB can achieve a similar saturation throughput as EB but with much smaller delay jitter. However, to the authors' best knowledge, no analysis was provided to explain the root cause behind the phenomenon.

To fully address the important questions $\textbf{Q1-Q3}$, this paper attempts to uncover the fundamental laws that govern the throughput stability and tail distribution of medium access delay. Our main contributions are detailed below.
\begin{itemize}
  \item[\textbf{C1}:] We find that the heaviness of the tail distribution of medium access delay is closely related to how rapidly the contention window is augmented with each collision. Specifically, EB always induces power law delay distribution regardless of the choice of backoff exponent $r$. Meanwhile, power law delay is mitigated as long as the backoff function is slower than exponential functions, i.e., $g(k)\in o\left(r^k\right)$ for all $r>1$, where $o\left(\cdot\right)$ will be defined more rigorously later. This explains the observations made by \cite{2008:Xu} and \cite{2011:Sun}. Furthermore, we find that delay distribution becomes light-tailed if the backoff function increases linearly or sub-linearly.
  \item[\textbf{C2}:] We prove that throughput stability is achieved only when the backoff function is at least as fast as an exponential function, i.e., $g(k)\in\Omega\left(r^k\right)$ for some $r>1$, where $\Omega(\cdot)$ will be defined rigorously later. In other words, PB fails to sustain non-zero asymptotic throughput, although they yield convergent delay moments. This presents a fundamental tradeoff between throughput stability and the heaviness of tail delay distribution.
  \item[\textbf{C3}:] We find that super-linear polynomial backoff achieves high throughput across a wide range of practical network size, despite its throughput instability asymptotically. This, together with our findings in $\textbf{C1}$, suggests that super-linear polynomial backoff is a better alternative than EB in supporting broadband network applications that call for both high throughput and low delay and delay jitter.
  \end{itemize}

Our study on the delay tail distribution of backoff process is not only for theoretical interest but also closely related to engineering applications. In the past few years, a number of modified exponential backoff schemes, including quality of service enhancing protocols, have been proposed to improve the delay performance of conventional BEB \cite{2004:Deng,2008:Barcelo,2003:Mangold}. For instance, \cite{2004:Deng} proposed a LMILD backoff algorithm, in which the contention window doubles upon collisions whereas decreases linearly upon successful transmissions. Besides, the enhanced distributed channel access (EDCA) scheme, which is adopted in the $802.11e$ standard, gives priority to delay-sensitive applications by setting a shorter contention window and shorter arbitration inter-frame space \cite{2003:Mangold}. Despite their respective contributions, they do not eliminate the fundamental feature of power law delay distribution induced by exponential backoff, and thus may still perceive relatively large delay jitter or high packet loss rate. Instead, we propose to fundamentally solve the power law delay problem by replacing EB with PB. Meanwhile, we show that high throughput can be achieved in a wide range of practical network size through parameter tuning of PB. In this sense, we can mitigate the power law delay distribution of EB without hurting the advantageous throughput performance. Our simulation results show that PB with reasonable backoff parameter outperforms BEB regardless of the existence of the retry limit. With current hardware processing power, the implementation of PB in random access networks incurs minor extra cost. Therefore, we believe it is a promising algorithm with broad applications in future random access networks.

The rest of the paper is organized as follows. We briefly review the backoff protocols and introduce some background information in Section II. In Section III, the main results of this paper is summarized. In Section IV, we analyze the power law tail distribution of medium access delay for general backoff protocols. In Section V, we derive the condition to sustain stable throughput. Simulation results are presented in Section VI, where we show that PB is a better alternative than EB in random access networks. Finally, the paper is concluded in Section VII.

\section{System Model and Preliminaries}
In this section, we first briefly review the operation of general backoff protocols. We then introduce the notion of medium access delay and some important metrics that will be used in later sections to evaluate the performance of different backoff schemes.
\begin{figure}
\centering
  \begin{center}
    \includegraphics[width=3.5in]{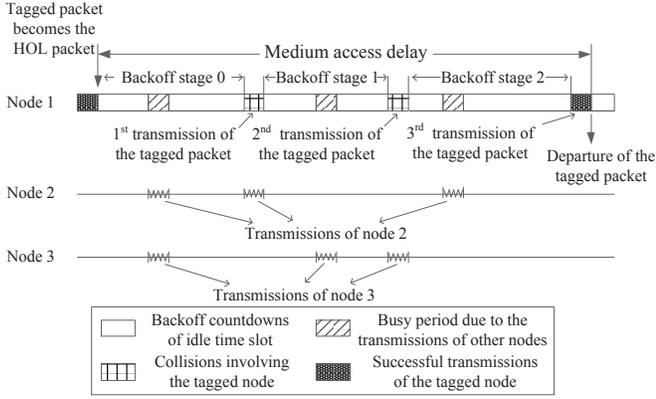}
  \end{center}
  \caption{Backoff process and medium access delay.}
  \label{52}
\end{figure}

\subsection{Backoff protocol operations}
We consider a fully connected WLAN consisting of $N$ continuously backlogged nodes. Illustrated in Fig. $\ref{52}$, the transmission of $3$ nodes is coordinated by a backoff mechanism. At each packet transmission, a node sets a backoff counter value $B$ by randomly choosing an integer from a contention window $[0, W-1]$, where $W$ is the size of contention window. At the initial transmission attempt of a packet, $W$ is set to its minimum value $W_0$. The contention window size is incremented on each collision. After $k^{th}$ collision, we say the node is in its $k^{th}$ backoff stage and the contention window size $W_k=g(k)W_0$, where $g(k)$ is an increasing backoff function characterizing the backoff process. For example, $g(k)=r^k$, $r > 1$ for EB, $g(k)=r^{k^a}$, $r > 1$ and $a<1$ for sub-exponential backoff (SEB), $g(k)=1+ k^b$, $b>0$ for polynomial backoff (PB). We denote the backoff counter value at the $k^{th}$ backoff stage by $B_k$. The backoff counter value decreases by one following each time slot, which could either be an idle slot or a transmission time slot. The packet is transmitted once the backoff counter reaches zero. When there is a finite retry limit $K$, a packet is dropped if it has not been successfully transmitted after $K$ retransmissions. The backoff process is illustrated in Fig. $\ref{52}$. For example, a tagged packet at node $1$ experiences three backoff stages before a successful transmission. In backoff stage $0$, one backoff countdown slot is occupied by the collision between node $2$ and $3$, while the other countdown slots are idle time slots. After the countdown process in stage $0$, node $1$ collides with node $2$'s transmission and enters backoff stage $1$. The process repeats until it successfully transmits after backoff stage $2$.

\begin{figure}
\centering
  \begin{center}
    \includegraphics[width=3.5in]{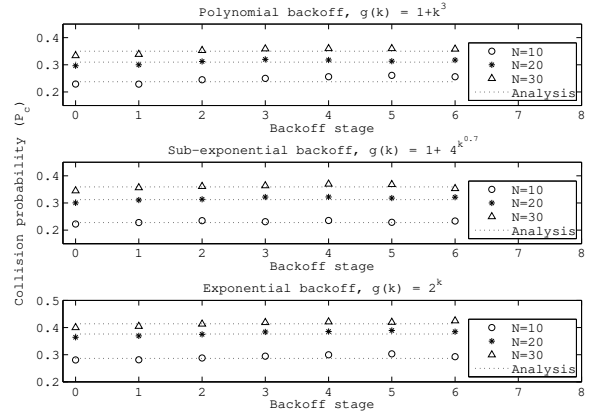}
  \end{center}
  \caption{Collision probabilities against backoff stage for different network size $N$ and backoff schemes. The initial backoff window $W_0=32$ and the retry limit $K=\infty$.}
  \label{60}
\end{figure}

The exact backoff process is very complex and often intractable, since the success and collision processes of various nodes are coupled and strongly correlated \cite{2008:Benaim}. A common technique adopted by most of the prior work on saturation analysis is the mean field decoupling approximation, where the backoff process at one node is decoupled and treated as if it is independent from the backoff processes at the other nodes. Specifically, it assumes that a node encounters a collision probability $P_c$ when it transmits, regardless of its own backoff stage. Moreover, the average attempt rate of an arbitrary node in a generic time slot, denoted by $\tau$, is assumed to be constant and does not vary with the backoff stage \cite{2000:Bianchi}. The validity of mean field approximation for EB has been recently verified in both theorem and experiments \cite{2009:Cho,2010:Huang}. To validate the assumption for general backoff functions, we reproduce the experiments in \cite{2010:Huang} and plot $P_c$ against backoff stage for three representative backoff functions in Fig. $\ref{60}$. The figure shows that, $P_c$ is largely independent of the backoff stage for all the backoff functions in consideration. In particular, the variances of $P_c$ across backoff stage are less than $0.01$ for all three backoff functions. The simulation results match the analytical results based on mean field approximation, which will be introduced in (\ref{1}) and (\ref{3}). Therefore, we can safely adopt the assumptions in this paper to analyze the throughput and delay performance of general backoff schemes.

Under mean field approximation, the probabilities of a time slot being an idle time slot, successful transmission or a collision can be calculated as
\begin{equation}
\label{19}
\begin{aligned}
P_{idle}&=\left(1-\tau\right)^N,\\
P_{succ}&=N \tau \left(1-\tau\right)^{N-1},\\
P_{coll}&=1-P_{succ}-P_{idle}.
\end{aligned}
\end{equation}
With (\ref{19}), normalized throughput $S$, defined as the portion of time occupied by successful packet transmissions, is given by
\begin{equation}
\label{49}
S=\frac{P_{succ}T_{succ}}{P_{idle}T_{idle} + P_{succ}T_{succ} +P_{coll}T_{coll}},
\end{equation}
where $T_{idle}$, $T_{succ}$ and $T_{coll}$ denote the lengths of idle, success and collision time slots, respectively. An important metric of system performance is throughput stability. Here, we say a backoff scheme is throughput-stable if it can yield non-zero asymptotic throughput when the network size approaches infinity (i.e., becomes extraordinarily large).

It is shown in \cite{2005:Bianchi} that the transmission probability $\tau$ of a saturation network in steady-state is the root of a fixed point system
\begin{equation}
\label{2}
\tau=\frac{\sum_{k=0}^K P_c^k}{\sum_{k=0}^K P_c^k+\sum_{k=0}^K P_c^kE[B_k]},
\end{equation}
where $P_c$ is the probability of a node encounters a collision when it transmits, which is given by
\begin{equation}
\label{1}
P_c=1-(1-\tau)^{N-1}.
\end{equation}
The above fixed point system always has one unique solution as long as the backoff function $g(k)$ is non-decreasing for $k=0,1,..,K$ \cite{2007:Kumar}. As we will show in later sections, some important properties of system performance, such as power law delay behavior and asymptotic throughput, are closely related to the value of $P_c$. It is worth noting that, $P_c<1$ always holds in a system under steady-state, despite that the limit of $P_c$ could be $1$. Otherwise, if $P_c=1$, all nodes will continuously encounter collisions and enter the next backoff stage. In this case, the limiting distribution of backoff stages does not exist and the system can never be in steady-state.

For simplicity, we assume that the retry limit $K$ is infinite hereafter, so as to better understand the factors that fundamentally affect the properties of a backoff function without considering the implementation details. In this case, (\ref{2}) becomes
\begin{equation}
\label{3}
\tau=\frac{1}{1+\left(1-P_c\right)\sum_{k=0}^{\infty}P_c^k E[B_k]}.
\end{equation}
The results obtained from the infinite-$K$ model can be easily translated to the standard systems with finite $K$. For instance, an infinite-variance medium access delay distribution in the infinite-$K$ model indicates high packet loss rate in the standard networks. In fact, we will show this implication in the Simulations section where results for both cases are presented.

\subsection{Medium access delay}
Unless otherwise stated, we use ``delay" and ``medium access delay" interchangeably throughout the paper. Illustrated in Fig. $\ref{52}$, the medium access delay of a packet, denoted by $X$, is the time period from the instant it becomes the head-of-line (HOL) packet to the instant at which the packet is successfully transmitted. Medium access delay of a packet consists of three parts, namely a series of backoff countdowns, collisions involving the tagged node and successful transmissions of the tagged node\cite{2010:Zhang}. In particular, the backoff countdown slots seen by a tagged node include idle slots as well as the busy slots, successful or collided, due to other nodes. For example, the medium access delay of the tagged packet at node $1$ in Fig. $\ref{52}$ includes $3$ backoff countdown stages, $2$ collision slots involving node $1$ and a successful transmission slot.

Suppose that a packet is successfully transmitted after $j$ collisions. Then, the medium access delay of this packet denoted by $X_j$, is
\begin{equation}
\label{8}
\begin{aligned}
X_j = \sum_{k=0}^{j}C_k + jT_{coll} + T_{succ}.
\end{aligned}
\end{equation}
Here, $C_k$ is the time consumed on backoff countdown at the $k^{th}$ backoff stage. It is the summation of a number of backoff countdown slots, given by
\begin{equation}
\label{32}
 C_k=\sum_{m=1}^{B_k}L_m,
\end{equation}
where $L_m$ is the length of its $m^{th}$ countdown slot, which could either be $T_{coll}$, $T_{succ}$ or an idle time slot $T_{idle}$. We denote the probability density function of medium access delay $X$ by $f(x)$ and its tail distribution function $\overline{F}(x)=\int_{x}^{\infty}f(t)dt$.

\subsection{Power law and heavy-tailed distributions}
\begin{figure}
\centering
  \begin{center}
    \includegraphics[width=3.5in]{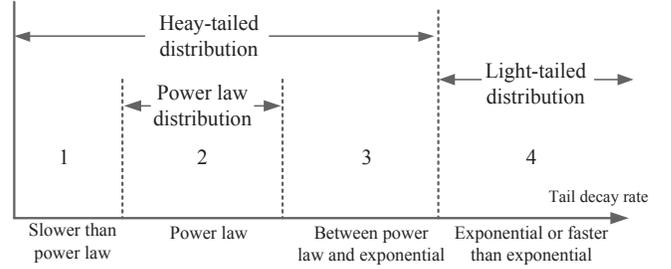}
  \end{center}
  \caption{Illustration tail distributions with respect to tail decay rate.}
  \label{53}
\end{figure}

In many engineering applications, we often encounter heavy-tailed distributions whose tail decaying rate is slower than exponential\cite{2003:Asmussen}. For instance, it is observed that both the size of data files stored in web servers and the process execution time in a computing environment follow heavy-tailed distribution\cite{2001:Crovella}. Power law distribution belongs to a subclass of heavy-tailed distribution, whose tail distribution follows a power law decaying rate. Conversely, a probability distribution is light-tailed if it is not heavy-tailed. The definitions of the different distributions with respect to the decay rate of tail distribution are formally defined as follows and illustrated in Fig. $\ref{53}$.

\textbf{Definition 1:} A probability distribution $f(x)$ is power law distribution with slope parameter $\alpha$, if its tail distribution function $\overline{F}(x)$ satisfies
\begin{equation}
\label{44}
\overline{F}(x) \sim x^{-\alpha}\mathcal{L}(x),
\end{equation}
where $\mathcal{L}(x)$ is a slow varying function (i.e., slower than any power function, such that $\lim_{x\rightarrow \infty} \frac{\mathcal{L}(tx)}{\mathcal{L}(x)}=1$ for all $t>0$. For instance, $\mathcal{L}(x)=\log(x)$) and the notation $h(x)\sim g(x)$ means $\lim_{x\rightarrow \infty} h(x)/g(x)=1$.

\textbf{Remark 1:} Power law distribution can also be characterized by the moments of $X$. For a power law distribution with slope parameter $\alpha$, $E\left[X^n\right]$ is finite for all $n<\alpha$ and is infinite for all $n\geq\alpha$ \cite{2009:Cho}. In fact, the tail decaying rate of a probability distribution is closely related to the convergence of moments. Specifically, a finite $E\left[X^n\right]$ indicates that the tail distribution decays faster than a polynomial function with power $n$ (cf. \cite{2005:Gut}, p. $75$). If $E\left[X^n\right]$ are finite for all $n\in \mathbb{N}$, the tail distribution of $f(x)$ decays faster than all power law functions and $f(x)$ belongs to region $3$ or $4$ in Fig. $\ref{53}$. In this case, we say that the power law distribution is mitigated.

\textbf{Definition 2:} A probability distribution $f(x)$ is heavy-tailed distribution if its moment generating function diverges, i.e.,
\begin{equation}
\label{21}
\int_0^{\infty} e^{\lambda x}f(x)dx=\infty,\ \  \forall \lambda>0.
\end{equation}

Using Taylor expansion to (\ref{21}), it holds that
\begin{equation}
\label{26}
\int_0^{\infty} e^{\lambda x}f(x)dx = \sum_{n=0}^{\infty} \frac{\lambda^n}{n!}\int_{0}^{\infty} x^n f(x) dx = \sum_{n=0}^{\infty} \frac{\lambda^n}{n!}\text{E}\left[X^n\right].
\end{equation}

\textbf{Remark 2:} The tail decay rate of a heavy-tailed distribution is slower than any exponential functions. From (\ref{26}), any divergent moment $\text{E}\left[X^n\right]$ would indicate that $f(x)$ is heavy-tailed distribution, but not the reverse. The RHS of (\ref{26}) could still be infinite even if all moments are finite. For example, Weibull distribution with shape parameter smaller than $1$ is a heavy-tailed distribution but not a power law distribution. The relationship between power law distribution and heavy-tailed distribution is shown in Fig. $\ref{53}$. Weibull distribution with shape parameter smaller than $1$ belongs to the set of distributions in region $3$ in Fig. $\ref{53}$.

Generally speaking, if a delay distribution is identified as a power law or heavy-tailed distribution, the probability of extremely large delay occurs is non-negligible. For instance, it is shown in \cite{2009:Cho} and \cite{2007:Sakurai} that the delay distribution of EB follows power law distribution, which is considered as the root cause of poor delay performance and user unfairness in current WLAN systems. In this paper, we develop a unified framework to study the heaviness of delay tail distribution of any general backoff functions, making the study in \cite{2009:Cho} and \cite{2007:Sakurai} a special case of ours.

Before leaving this session, we introduce the following two important notations to describe the limiting behavior of functions.

\textbf{Definition 3:} A function $g(x)\in \Omega(f(x))$, if $\exists c>0$ and $\exists x_0$ such that $g(x)\geq cf(x)$, $\forall x>x_0$.

\textbf{Definition 4:} A function $g(x)\in o(f(x))$, if $\forall c>0$, $\exists x_0$ such that $|g(x)|\leq c|f(x)|$, $\forall x>x_0$.

\textbf{Remark 3:} Loosely speaking, $g(x)$ is asymptotically ``no slower" than $f(x)$ if $g(x) \in \Omega\left(f(x)\right)$ and ``slower" than $f(x)$ if $g(x) \in o\left(f(x)\right)$. The definitions of $\Omega(\cdot)$ and $o(\cdot)$ can be straightforwardly extended to discrete functions $g(k)$ and $f(k)$ with the replacements of $x$ by $k$ and $x_0$ by $k_0$, where $k,k_0\in\mathbb{N}$. In the strict sense, $g(k)\notin \Omega\left(f(k)\right)$ does not imply $g(k)\in o\left(f(k)\right)$, and vice versa. However, under a mild condition that the limit of $\lim_{k\rightarrow \infty} \frac{g(k)}{f(k)} = L$ exists ($0 \leq L\leq \infty$), the two notations $g(k)\in \Omega\left(f(k)\right)$ and $g(k)\in o\left(f(k)\right)$ are complementary (cf. \cite{2009:Cormen}, Ch. 3). Specifically, $g(k)\in o\left(f(k)\right)$ if $L=0$ and $g(k)\in \Omega\left(f(k)\right)$ otherwise. A special interest of this paper is to compare the growth rate of a general backoff function $g(k)$ with an exponential function, i.e. $f(k)=r^k$. In this case, the limit $\lim_{k\rightarrow \infty} \frac{g(k)}{f(k)}$ exists for most of the practical backoff functions, such as EB, SEB and PB. Without causing confusions, we discuss in the following in the weaker sense that $g(k)\in \Omega(r^k)$ if and only if $g(k)\in o(r^k)$ fails.

\section{Main Results}
We summarize in this section the key results of this paper. The proofs of the results are deferred to Section IV and V.
\begin{itemize}
  \item \textbf{Power law delay}: A random-access network with an increasing backoff function $g(k)$ does not suffer a power law delay if and only if $g(k)\in o\left(r^k\right)$, $\forall r>1$. In other words, the system observes power law delay if and only if $\exists r>1$ such that $g(k)\in\Omega(r^k)$ (proved in Section IV).
    \item \textbf{Heavy-tailed delay}: The distribution of medium access delay is heavy-tailed with EB, SEB and superlinear PB, while light-tailed with linear-sublinear PB (proved in Appendix B).
  \item \textbf{Throughput stability}: An increasing backoff function $g(k)$ is throughput-stable if and only if $\exists r>1$ such that $g(k)\in\Omega\left(r^k\right)$. In other words, the network is throughput-unstable if and only if $g(k)\in o\left(r^k\right)$, $\forall r>1$ (proved in Section V).
\end{itemize}

Our results show that the power law behavior of EB is essentially attributed to its exponential function growth rate, and can be mitigated if EB is replaced by a ``slower" backoff function, such as PB and SEB. However, it is also the exponential growth rate that ensures non-zero asymptotic throughput of EB when the network size becomes extraordinarily large. This also implies that it is impossible to achieve stable throughput and non-power law medium access delay distribution at the same time.

\begin{figure}
\centering
  \begin{center}
    \includegraphics[width=3.5in]{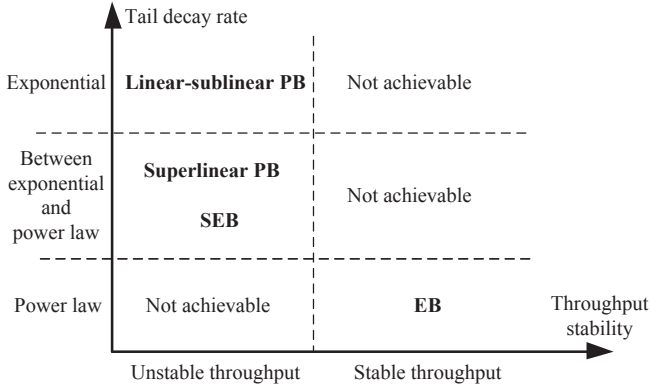}
  \end{center}
  \caption{Operation regions of different backoff functions.}
  \label{54}
\end{figure}

Using PB, SEB and EB as examples, Fig. $\ref{54}$ summarizes the key results of this paper. It shows that the heaviness of tail distribution improves from a power law tail with EB, to a heavy but non-power law tail with SEB and superlinear PB, and eventually to a light tail with linear-sublinear PB. However, stable throughput is unattainable when EB is replaced by the other backoff schemes.

\section{Analysis of Power Law Behavior of Medium Access Delay}
In this section, we characterize the power law behavior of medium access delay distribution for general backoff functions. This is achieved by studying the convergence of moments of medium access delay. We prove that convergent delay moments are attainable if and only if the backoff functions are ``slower" than exponential function. Accordingly, backoff functions such as EB and $g(k)=r^{k^a}$ with $r,a\geq 1$, always induce power law delay distribution. In contrast, PB and SEB can fully eliminate the power law tail of medium access delay.

\subsection{Moments of medium access delay}
Following the definition of $P_c$, the probability that a packet is successfully transmitted after $j$ consecutive collisions is $P_c^j(1-P_c)$. Therefore, the $n^{th}$ ($n\in \mathbb{N}$) moment of the medium access delay $X$ is
\begin{equation}
\label{9}
\begin{aligned}
&E\left[X^n\right]\\
=&(1-P_c) \sum_{j=0}^{\infty}P_c^jE\left[X_j^n\right] \\
=&(1-P_c) \sum_{j=0}^{\infty}P_c^jE\left[\left\{\sum_{k=0}^{j}C_k + jT_{coll} + T_{succ}\right\}^n\right]\\
=&(1-P_c)\sum_{j=0}^{\infty}P_c^jE\left[\left(\sum_{k=0}^{j}C_k\right)^n + \right.\\
 &\left.\text{ other terms with power of } \sum_{k=0}^{j}C_k \text{ lower than } n-1\right],
\end{aligned}
\end{equation}
where $C_k$ is the time consumed on backoff countdown at the $k^{th}$ backoff stage, given in (\ref{32}). We can see that the convergence of $E\left[X^n\right]$ is determined by the most significant term in the RHS of (\ref{9}), i.e.
\begin{equation}
\label{10}
\begin{aligned}
(1-P_c)\sum_{j=0}^{\infty}P_c^jE\left[\left(\sum_{k=0}^{j}C_k\right)^n\right].
\end{aligned}
\end{equation}
That is to say, the convergence of medium access delay is equivalent to that of the integrated backoff countdown process.

Similarly, let $\Lambda$ denote the total number of backoff countdowns before the packet successfully transmits. If a packet is successfully transmitted after $j$ collisions, the total number of backoff countdowns denoted by $\Lambda_j$, is
\begin{equation}
\begin{aligned}
\Lambda_j = \sum_{k=0}^j B_k,
\end{aligned}
\end{equation}
where $B_k$ is the backoff counter value at the $k^{th}$ backoff stage. Therefore, the $n^{th}$ moment of $\Lambda$ is
\begin{equation}
\label{11}
\begin{aligned}
E\left[\Lambda^n\right]=(1-P_c)\sum_{j=0}^{\infty}P_c^j E\left[\left(\sum_{k=0}^j B_k\right)^n\right].
\end{aligned}
\end{equation}
Recall that $ C_k=\sum_{m=1}^{B_k}L_m$. Besides, the countdown time slot $L_m$ is bounded as
\begin{equation*}
\min\left(T_{idle}, T_{succ},T_{coll}\right)\leq L_m \leq \max\left(T_{idle}, T_{succ},T_{coll}\right).
\end{equation*}
Therefore, (\ref{10}) is lower bounded by
\begin{equation}
\label{16}
\left\{\min\left(T_{idle}, T_{succ},T_{coll}\right)\right\}^n\cdot E\left[\Lambda^n\right]
\end{equation}
meanwhile upper bounded by
\begin{equation}
\left\{\max\left(T_{idle}, T_{succ},T_{coll}\right)\right\}^n\cdot E\left[\Lambda^n\right].
\end{equation}
It can be seen that (\ref{10}) converges if and only if $E\left[\Lambda^n\right]$ converges. That is to say, the convergence of $E\left[X^n\right]$ is equivalent to that of $E\left[\Lambda^n\right]$. In this sense, we mainly focus on the convergence properties of $E\left[\Lambda^n\right]$ in the following discussions.

\subsection{Power law delay analysis}
Theorem $1$ presents the relation between power law delay distribution and backoff function growth rate.

\textbf{Theorem 1:} A random-access network with an increasing backoff function $g(k)$ suffers a power law delay if $\exists r>1$ such that $g(k)\in\Omega(r^k)$, and does not suffer a power law delay if $g(k)\in o\left(r^k\right)$, $\forall r>1$.

\emph{Proof}: We first prove that a $g(k)$ suffers a power law delay if $\exists r>1$ such that $g(k)\in\Omega(r^k)$. To prove the argument, we only need to show that there exists an infinite $E\left[\Lambda^n\right]$.

By the Jensen's inequality,
\begin{equation}
\label{33}
E\left[\left(\sum_{k=0}^j B_k\right)^n\right] \geq \left[E\left(\sum_{k=0}^{j}B_k\right)\right]^n = \left(\sum_{k=0}^{j}E\left[B_k\right]\right)^n,
\end{equation}
for $\forall n\in \mathbb{N}$.
Substituting (\ref{33}) into (\ref{11}), we have the lower bound of $E[\Lambda^n]$, where
\begin{equation}
\label{22}
\begin{aligned}
E[\Lambda^n]&\geq\left(1-P_c\right)\sum_{j=0}^{\infty}P_c^j \left(\sum_{k=0}^{j}E\left[B_k\right]\right)^n\\
&\geq \left(1-P_c\right)\sum_{j=0}^{\infty}\left\{P_c^j \sum_{k=0}^{j}\left(E\left[B_k\right]\right)^n\right\}\\
&=\frac{1}{2^n}\left(1-P_c\right)\sum_{j=0}^{\infty}\left\{P_c^j \sum_{k=0}^{j}\left(W_k-1\right)^n\right\}.
\end{aligned}
\end{equation}
The last equality holds because $E\left[B_k\right]=\frac{W_k-1}{2}$.

For a $g(k)\in\Omega(r^k)$, there always $\exists c>0$ and $\exists k_0$ such that $g(k)\geq c r^k+\frac{1}{W_0}$, $\forall k>k_0$. With $W_k=g(k)W_0$, the RHS of (\ref{22}) is be lower bounded by
\begin{equation}
\label{36}
\begin{aligned}
&\frac{1}{2^n}\left(1-P_c\right)\sum_{j=k_0+1}^{\infty}\left\{P_c^j\sum_{k=0}^{j}\left(W_k-1\right)^n\right\}\\
\geq &\frac{1}{2^n}\left(1-P_c\right)\sum_{j=k_0+1}^{\infty}\left\{P_c^j \sum_{k=k_0+1}^{j}\left(W_k-1\right)^n \right\}\\
\geq & \left(\frac{cW_0}{2}\right)^n\left(1-P_c\right) \sum_{j=k_0+1}^{\infty} \left\{P_c^j \sum_{k=k_0+1}^{j}r^{kn} \right\}\\
= & \left(\frac{cW_0}{2}\right)^n\frac{r^{(k_0+1)n}\left(1-P_c\right)}{r^n-1}\sum_{j=k_0+1}^{\infty} \left\{\frac{\left(P_cr^n\right)^j}{r^{nk_0}} - P_c^j\right\}.
\end{aligned}
\end{equation}
Notice that the above lower bound of $E[\Lambda^n]$ becomes infinite when $P_cr^n\geq 1$, or equivalently $n\geq -\frac{\ln P_c}{\ln r}$. This leads to the proof that $g(k)$ suffers a power law delay

Then, we prove that $g(k)$ does not suffer a power law delay if $g(k)\in o\left(r^k\right)$, $\forall r>1$. This is equivalent to show that $E\left[\Lambda^n\right]$ is finite for all $n\in \mathbb{N}$. According to the Holder's inequality, it holds that
\begin{equation}
\left(\sum_{k=0}^j B_k\right)^n\leq \left(j+1\right)^{n-1}\sum_{k=0}^jB_k^n.
\end{equation}
Taking the expectations on the both sides, we have
\begin{equation}
E\left[\left(\sum_{k=0}^j B_k\right)^n\right]\leq E\left[\left(j+1\right)^{n-1}\sum_{k=0}^{j}B_k^n \right].
\end{equation}
From (\ref{11}), $E[\Lambda^n]$ is upper bounded by
\begin{equation}
\label{12}
\begin{aligned}
E[\Lambda^n]&\leq\left(1-P_c\right)\sum_{j=0}^{\infty}\left\{P_c^j\left(j+1\right)^{n-1} \sum_{k=0}^{j}E\left[B_k^n \right]\right\}.
\end{aligned}
\end{equation}
By assumption, $B_k$ is uniformly generated from $\left[0,W_k-1\right]$. Thus, we have
\begin{equation}
\begin{aligned}
E\left[B_k^n \right]&=\frac{1}{W_k}\cdot \sum_{l=0}^{W_k-1} l^n \leq \frac{1}{W_k} \int_{0}^{W_k}x^n\ dx = \frac{1}{n+1}W_k^{n}.
\end{aligned}
\end{equation}
Substituting $E\left[B_k^n \right]\leq \frac{1}{n+1}W_k^{n}$ into (\ref{12}), we have
\begin{equation}
\label{74}
E[\Lambda^n]\leq\frac{1-P_c}{n+1}\sum_{j=0}^{\infty}\left\{P_c^j\left(j+1\right)^{n-1}\cdot \sum_{k=0}^{j}W_k^{n}\right\}.
\end{equation}

By definition, for any $r>1$ and $c>0$ there exists a $k_r>0$, such that $g(k)\leq cr^k$ for all $k>k_r$. Then, the following inequality holds for for all $r>1$ and $c>0$,
\begin{equation}
\label{61}
\begin{aligned}
&E[\Lambda^n]\leq\frac{\left(1-P_c\right)W_0^n}{n+1}\Biggl\{\sum_{j=0}^{k_r}\left[P_c^j\left(j+1\right)^{n-1}\cdot \sum_{k=0}^{j}g^n(k)\right]+ \\
&\sum_{j=k_r+1}^{\infty}\left[P_c^j\left(j+1\right)^{n-1} \left(\sum_{k=0}^{k_r}g^n(k) + \sum_{k=k_r+1}^{j}c^nr^{nk}\right)\right]\Biggr\}.
\end{aligned}
\end{equation}
The second term in the RHS of (\ref{61}), which determines the convergence of the upper bound, can be expressed as
\begin{equation}
\label{62}
\begin{aligned}
\sum_{j=k_r+1}^{\infty}\left\{\mu_r P_c^j\left(j+1\right)^{n-1}+ \frac{\left(cr\right)^n}{r^n-1}\left(P_cr^n\right)^j\left(j+1\right)^{n-1}\right\},
\end{aligned}
\end{equation}
where
\begin{equation}
\mu_r = \sum_{k=0}^{k_r}g^n(k) - \frac{c^nr^{n\left(k_r+1\right)}}{r^n-1}
\end{equation}
is a finite constant for a given $r$. Noticeably, (\ref{62}) is the upper bound on $E\left[\Lambda^n\right]$ for any $r>1$. Thus, we can safely say $E\left[\Lambda^n\right]$ is finite as long as there exists an $r>1$ such that (\ref{62}) is finite.

Notice that the first term in (\ref{62})
\begin{equation}
\begin{aligned}
\sum_{j=k_r+1}^{\infty}\mu_r P_c^j\left(j+1\right)^{n-1}
\end{aligned}
\end{equation}
is convergent for all $n\in \mathbb{N}$. Then, (\ref{62}) is finite if and only if
\begin{equation}
\label{63}
\begin{aligned}
\sum_{j=k_r+1}^{\infty}\left(P_cr^n\right)^j\left(j+1\right)^{n-1}
\end{aligned}
\end{equation}
is finite. This can be achieved by selecting a
\begin{equation}
\label{65}
1<r<\left(\frac{1}{P_c}\right)^{\frac{1}{n}}.
\end{equation}
In other words, $\forall n \in \mathbb{N}$ we can always find a finite upper bound of $E[\Lambda^n]$ by selecting a $r$ as in (\ref{65}). This implies that all the delay moments are finite with the backoff function $g(k)$, i.e. power law delay is mitigated. $\hfill \blacksquare$

\textbf{Remark 4:} Loosely speaking, Theorem $1$ implies that a backoff function $g(k)$ will induce power law delay distribution if and only if $g(k)\in \Omega\left(r^k\right)$ for some $r>1$, i.e., $g(k)$ grows at least as fast as an exponential function. Therefore, EB will always induce power law delay distribution, while PB and SEB can mitigate the power law delay. This also explains the observations made by \cite{2008:Xu} and \cite{2011:Sun} that PB achieves better delay performance than EB.

\textbf{Corollary 1:} For an increasing backoff function $g(k)$, if $\lim_{j\rightarrow \infty} \frac{g(j+1)}{g(j)}$ exists and is denoted by
\begin{equation}
\label{78}
\lim_{j\rightarrow \infty} \frac{g(j+1)}{g(j)} = \gamma,
\end{equation}
where $1\leq \gamma<\infty$. Then, $E\left[\Lambda^n\right]$ is finite if and only if $P_c \gamma^n <1$. This implies that $g(k)$ yields a power law delay if $\gamma>1$, and a non-power law delay if $\gamma=1$.

\emph{Proof:} See Appendix A.  $\hfill \blacksquare$

The backoff functions discussed in Corollary $1$ are special cases of the general ones discussed in Theorem $1$, in that the limit of $\frac{g(j+1)}{g(j)}$ exists as $j\rightarrow \infty$. It is easy to check that $\gamma=r>1$ for EB ($g(k)=r^k$) and $\gamma=\lim_{j\rightarrow \infty} \frac{1+(j+1)^b}{1+j^b} = 1$ for PB ($g(k)=1+k^b$, $b>0$) and $\gamma = \lim_{j\rightarrow \infty} r^{(j+1)^a-j^a} = 1$ for SEB ($g(k)= r^{k^a}$, $0<a<1$). Thus, EB suffers a power law delay, while PB and SEB do not. This is consistent with Theorem $1$. Moreover, following Corollary $1$, we see that $E[X^n]$ for EB is finite if $P_c < \frac{1}{r^n}$ and infinite if $P_c \geq \frac{1}{r^n}$. This is consistent with the results in \cite{2009:Cho} and \cite{2007:Sakurai}. Notice that Theorem $1$ is more general in that it applies even if the limit in the LHS of (\ref{78}) does not exist \footnote{Consider an increasing function $g(k)$ with the even entries equal to $2^k$ and the odd entries equal to $3\cdot 2^{k-1}$. In this case, $g(k)\in\Omega\left(2^k\right)$ but $\lim_{j\rightarrow \infty} \frac{g(j+1)}{g(j)}$ does not exist.}.

Before leaving the section, note that a non-power law distribution could still be heavy-tailed, such as Weibull distribution with slope parameter smaller than $1$. It is of mathematical interest to further investigate the heavy-tailed behavior of delay distribution when different backoff functions are used, although the study may be of little practical relevance. For the completeness of mathematics, we provide detailed analysis on heavy tailed delay distribution in the Appendix B. Interestingly, our results show that, although SEB and super-linear PB ($g(k)=1+k^b,b>1$) can mitigate the power law tail distribution, the delay distributions are still heavy-tailed. Meanwhile, as we further decrease the growth rate of backoff function, delay distribution eventually becomes light-tailed when a linear-sublinear PB is used ($g(k)=1+k^b,b\leq1$). A point to mention is that, our analysis of heavy-tailed behavior of delay distribution is mainly for theoretical interest instead of engineering applications. On one hand, linear-sublinear PB is impractical in the sense that it may yield prohibitively low throughput. On the other hand, although the delay distributions are heavy-tailed with SEB and super-linear PB, their tail decaying rates are essentially faster than all power law functions, which is sufficient for any practical engineering implementations. In fact, we will show the superior delay performance of SEB and PB in Simulations section.

\section{Stability of Saturation Throughput}
In this section, we show that stable throughput is attainable if and only if the backoff function grows at least as fast as an exponential function. Here, we say a backoff scheme is throughput-stable if it yields non-zero asymptotic throughput when the network size approaches infinity (i.e., becomes extraordinarily large). In other words, a throughput-stable backoff scheme guarantees non-zero throughput regardless of the network size, and thus is suitable for practical deployment where the network size can vary randomly over time.

\subsection{Asymptotic throughput analysis}
The collision probability $P_c$ increases as the network size $N$ increases due to higher contention level. Meanwhile, $\tau$ decreases with $N$, since the increased contention window size results in a smaller probability of transmission in a given time slot \cite{2005:Kwok}. Since $P_c<1$ and monotonically increases with $N$, the limit $\lim_{N\rightarrow \infty}P_c$ exists. Similar argument applies to $\tau$, such that $\lim_{N\rightarrow \infty}\tau$ exists as well. Besides, as $N$ approaches infinity, it holds that $\tau\rightarrow 0$ because almost all the nodes are in extremely high backoff stage when the network is stable, such that the probability of transmission approaches zero (to be justified in (\ref{68})).

Taking limit on both sides of (\ref{1}), we have
\begin{equation}
\label{46}
\begin{aligned}
\lim_{N\rightarrow \infty}1-P_c =&  \lim_{N\rightarrow \infty}\left(1-\tau\right)^{N-1}\\
=& \lim_{N\rightarrow \infty}\left(1-\tau\right)^{\frac{1}{\tau}(N-1)\tau} =\lim_{N\rightarrow \infty} e^{-(N-1)\tau},
\end{aligned}
\end{equation}
where the last equality holds because $\lim_{N\rightarrow\infty}\tau=0$. For simple illustration, we assume $T_{idle}=T_{succ}=T_{coll}$. Consequently, the normalized throughout in (\ref{49}) becomes $S=P_{succ}$. Notice that the conclusions in this section apply even without this assumption. Taking logarithm on both sides of (\ref{46}), we have
\begin{equation}
\lim_{N\rightarrow \infty}N\tau =\lim_{N\rightarrow \infty}\ln\left(\frac{1}{1-P_c}\right).
\end{equation}
Then, the asymptotic $P_{succ}$ in (\ref{19}) is
\begin{equation}
\label{14}
\begin{aligned}
\lim_{N\rightarrow\infty} P_{succ}& =  \lim_{N\rightarrow\infty} N\tau(1-\tau)^{N-1} \\
=& \lim_{N\rightarrow\infty}\ln \left(\frac{1}{1-P_c}\right)(1-P_c),
\end{aligned}
\end{equation}
where the last equality holds because $\left(1-\tau\right)^{N-1}=1-P_c$. For EB, it has been proved in \cite{2007:Kumar} that $\lim_{N \rightarrow \infty}P_c =\frac{1}{r}$. Evidently, its asymptotic throughput is non-zero following (\ref{14}). The throughput stability of general backoff functions is studied in the following subsection.

\subsection{Condition of stable throughput}
The following Theorem $2$ provides a criterion to determine the throughput stability of a general backoff function. Accordingly, we find that the throughput of SEB and PB collapses to zero when the network size is extraordinarily large.

\textbf{Theorem 2:} For a system in steady-state, an increasing backoff function $g(k)$ is throughput-stable if $\exists r>1$ such that $g(k)\in\Omega\left(r^k\right)$, and throughput-unstable if  $g(k)\in o\left(r^k\right)$ for all $r>1$.

\emph{Proof}:\ As per (\ref{2}) and (\ref{1}), the fixed point system for a backoff function $g(k)$ is
\begin{equation}
\label{72}
\tau=\frac{2}{1+W_0\left(1-P_c\right)\sum_{k=0}^{\infty} P_c^kg(k)} \triangleq \Theta\left(P_c\right)
\end{equation}
and
\begin{equation}
\label{73}
P_c = 1- (1-\tau)^{N-1} \triangleq \Phi\left(\tau\right).
\end{equation}
The limit of $\tau$ is
\begin{equation}
\label{68}
\lim_{N\rightarrow\infty}\tau = 1 - \lim_{N\rightarrow \infty}\left(1-P_c\right)^{\frac{1}{N-1}} = 0,
\end{equation}
as $P_c<1$ \footnote{Recall that $P_c<1$ strictly holds in a stable network, although the limit of $P_c$ could be $1$.}. From (\ref{72}), this indicates that $\lim_{N\rightarrow \infty}\sum_{k=0}^{\infty} P_c^kg(k)= \infty$.

\begin{figure}
\centering
  \begin{center}
    \includegraphics[width=2.8in]{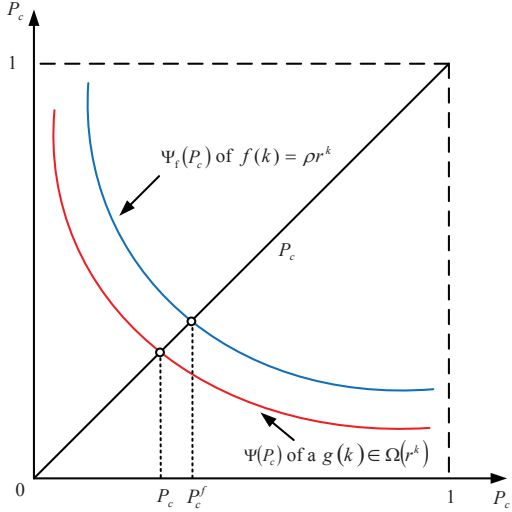}
  \end{center}
  \caption{Illustration of $\Psi_f\left(P_c\right)$, $\Psi\left(P_c\right)$ and the fixed point solutions for backoff functions $f(k)=\rho r^k$ and a $g(k)\in \Omega\left(r^k\right)$.}
  \label{79}
\end{figure}

Meanwhile, the fixed point system in (\ref{72}) and (\ref{73}) can be compactly written as $P_c=\Psi\left(P_c\right) \triangleq \Phi\left(\Theta\left(P_c\right)\right)$. Here, $\Psi\left(P_c\right)$ is an decreasing function in $P_c$. This can be justified by calculating
\begin{equation}
\begin{aligned}
\label{67}
&\frac{d \left\{\left(1-P_c\right)\sum_{k=0}^{\infty} P_c^kg(k)\right\}}{d P_c} \\
=& \sum_{k=0}^{\infty} \left(k+1\right)P_c^k \left[g(k+1)-g(k)\right]>0,
\end{aligned}
\end{equation}
where the inequality holds because $g(k)$ is an increasing function. As illustrated in Fig. $\ref{79}$, the solution of the fixed point system is the intersecting point of the $P_c$ and $\Psi\left(P_c\right)$ curves.

We first prove the first part of Theorem $2$. That is, the asymptotic throughput is strictly larger than zero if $g(k)\in\Omega\left(r^k\right)$ for some $r>1$. By definition, there $\exists k_0$ and $\exists c>0$ such that $g(k)\geq c r^k$, $\forall k\geq k_0$. Then, we have
\begin{equation}
\begin{aligned}
\sum_{k=0}^{\infty} P_c^kg(k) &= \sum_{k=0}^{k_0} P_c^kg(k) + \sum_{k=k_0+1}^{\infty} P_c^kg(k)\\
&\geq \sum_{k=0}^{k_0} P_c^kg(k) + \sum_{k=k_0+1}^{\infty}c \left(P_c r\right)^k\\
&= \sum_{k=0}^{k_0} P_c^kg(k) + \sum_{k=0}^{\infty} \left[c\left(P_cr\right)^{k_0+1}r^k\right] P_c^k.
\end{aligned}
\end{equation}
Let $\rho \triangleq c\left(P_cr\right)^{k_0+1}$ denote a constant parameter, we have
\begin{equation}
\label{64}
\begin{aligned}
\left(1-P_c\right)\sum_{k=0}^{\infty} P_c^kg(k) \geq \left(1-P_c\right)\sum_{k=0}^{\infty} P_c^k \left(\rho r^k\right).
\end{aligned}
\end{equation}

In the following, we show that $\lim_{N\rightarrow \infty} P_c\leq \frac{1}{r}$. This is proved by comparing $P_c$ resulting from $g(k)$ with that resulting from an exponential backoff function $f(k)=\rho \cdot r^k$. To distinguish from the notation $\Psi(P_c)$ for $g(k)$, we use $\Psi_f\left(P_c\right)$ to denote the fixed point system when $g(k)$ is replaced by $f(k)$. The corresponding fixed point solution is denoted by $P_c^f$. From (\ref{64}), we can infer that
\begin{equation}
\label{81}
\Psi_f\left(P_c\right)\geq \Psi\left(P_c\right),\ \  \forall P_c\in (0,1).
\end{equation}
We illustrate $\Psi_f\left(P_c\right)$ and $\Psi\left(P_c\right)$ in Fig. $\ref{79}$, where we observe that $P_c\leq P_c^f$. In fact, this can also be rigorously proved by contradiction. Assuming $P_c> P_c^f$, we have
\begin{equation}
\label{66}
P_c^f=\Psi_f\left(P_c^f\right) \geq \Psi_f\left(P_c\right) \geq \Psi\left(P_c\right)= P_c,
\end{equation}
where the first inequality holds because $\Psi_f\left(P_c\right)$ is an decreasing function of $P_c$ and the second inequality is from (\ref{81}). Clearly, (\ref{66}) leads to an contradiction to the assumption $P_c> P_c^f$. For EB with backoff function $\rho\cdot r^k$, $\lim_{N\rightarrow \infty}P_c^f =\frac{1}{r}$ regardless of the value of $\rho$ \footnote{This is because $\rho$ can be considered as a scaling factor to the initial contention window size $W_0$, which is unrelated to the asymptotic behavior of EB.}. Accordingly, we have $P_c\leq P_c^f\leq \frac{1}{r}$. Since $P_c$ is increasing with $N$, it suffices to conclude that $0<\lim_{N\rightarrow\infty}P_c\leq \frac{1}{r}$. In this case, the asymptotic throughput in (\ref{14}) is non-zero, indicating that $g(k)$ is a throughput-stable scheme.

Next, we prove the second part of Theorem $2$. That is, the asymptotic throughput is zero if $g(k)\in o\left(r^k\right)$ for all $r>1$. By definition, for any $r>1$ and $c>0$ there exists a $k_r>0$, such that $g(k)\leq cr^k$ for all $k>k_r$. For a given $r>1$, we have
\begin{equation}
\label{80}
\begin{aligned}
\lim_{N\rightarrow \infty}\sum_{k=0}^{\infty} P_c^kg(k) &= \sum_{k=0}^{k_r} P_c^kg(k) + \sum_{k=k_r+1}^{\infty} P_c^kg(k)\\
&\leq \sum_{k=0}^{k_r} P_c^kg(k) + c \sum_{k=k_r+1}^{\infty} \left(P_cr\right)^k.
\end{aligned}
\end{equation}
Since $\lim_{N\rightarrow \infty}\sum_{k=0}^{\infty} P_c^kg(k)= \infty$ and the inequality in (\ref{80}) holds for all $r>1$, we have
\begin{equation}
\begin{aligned}
\lim_{N\rightarrow \infty} \sum_{k=k_r+1}^{\infty} \left(P_cr\right)^k = \infty
\end{aligned}
\end{equation}
for all $r>1$. In other words, $\lim_{N\rightarrow \infty}P_cr\geq 1$ must hold for all $r>1$. This is achievable only if $\lim_{N\rightarrow \infty}P_c=\beta=1$, otherwise we can always find a $r<\frac{1}{\beta}$ such that $\lim_{N\rightarrow \infty}P_cr< 1$. With $\lim_{N\rightarrow \infty}P_c=1$, we can derive
\begin{equation}
\label{43}
\begin{aligned}
\lim_{N\rightarrow \infty}P_{succ}&=\lim_{P_c\rightarrow 1}\ln \left(\frac{1}{1-P_c}\right)(1-P_c) \\
&= \lim_{x\rightarrow \infty} \frac{\ln x}{x}=\lim_{x\rightarrow \infty}\frac{1}{x} = 0,
\end{aligned}
\end{equation}
indicating a zero asymptotic throughput. This completes the proof of the second part of Theorem $2$. $\hfill \blacksquare$

Theorem $2$ implies that a backoff scheme is throughput-stable if and only if the backoff function grows at least as fast as an exponential function. Accordingly, stable throughput is attainable with EB, but unattainable with PB or SEB. Evidently, there exists a tradeoff between the tail decaying rate of medium access delay and throughput stability. When ``faster" backoff function is used, the tail distribution of delay becomes heavier while throughput stability improves, and vice versa. In practice, we need to select backoff functions that achieve a balance of the tradeoff. Interestingly, we show in the Simulation section that PB can achieve good throughput and delay performance within practical range of user population, when the order of backoff function is set properly. We therefore advocate PB as a potential candidate to replace EB in current random access networks, now that there are increasingly more multimedia applications with stringent delay requirements in the network.

\section{Simulation Results}
In this section, we first verify our analysis in previous sections. We then illustrate through numerical simulations that superlinear PB is a good alternative of EB due to its good throughput performance and finite delay moments.

\begin{table}
\footnotesize
\caption{System Parameters of $802.11g$}
\begin{center}
\begin{tabular}{|l|l|}
\hline
 PHY layer transmission rate (R)          &   $54$ Mbps    \\ \hline
 PHY preamble \& header  ($P^h$)          &   $24$ $\mu s$  \\ \hline
 MAC header \& FCS    ($M^h$)             &   $272$ bits transmitted at $54$ Mbps    \\ \hline
 DIFS                                     &   $34$ $\mu s$    \\ \hline
 SIFS                                     &   $16$ $\mu s$    \\ \hline
 Mini slot time $\sigma$                  &   $9$ $\mu s$    \\ \hline
 ACK                                      &   $24.5$ $\mu s$           \\ \hline
 Pay Load (PL)                            &   $1500$ bytes ($12000$ bits)  \\ \hline
 $W_0$                                    &   $16$            \\ \hline
 Retry limit                              &   $\infty$           \\ \hline
\end{tabular}
\end{center}
\end{table}

\subsection{Validations of analytical results}
Unless otherwise stated, we use the DCF basic-access mode in $802.11g$, where the system parameters are listed in Table I. The slot lengths are
\begin{equation*}
\begin{cases}
T_{idle} = \sigma \\
T_{succ} = P^h + M^h + PL/R + SIFS + ACK + DIFS\\
T_{coll} = P^h + M^h + PL/R + DIFS.
\end{cases}
\end{equation*}

\begin{figure}
\centering
  \begin{center}
    \includegraphics[width=3.5in]{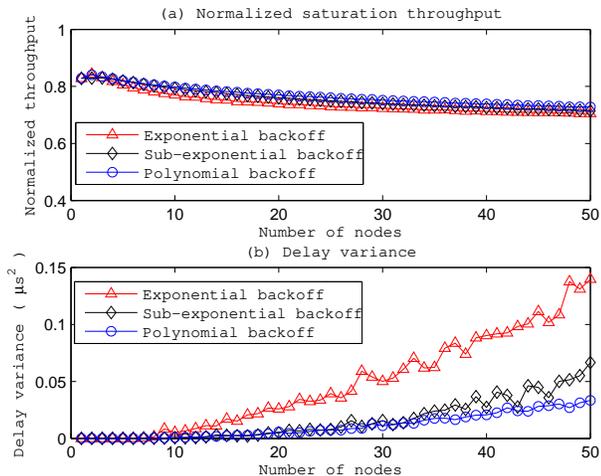}
  \end{center}
  \caption{Performance comparisons with retry limit $K=\infty$.}
  \label{55}
\end{figure}

\begin{figure}
\centering
  \begin{center}
    \includegraphics[width=3.5in]{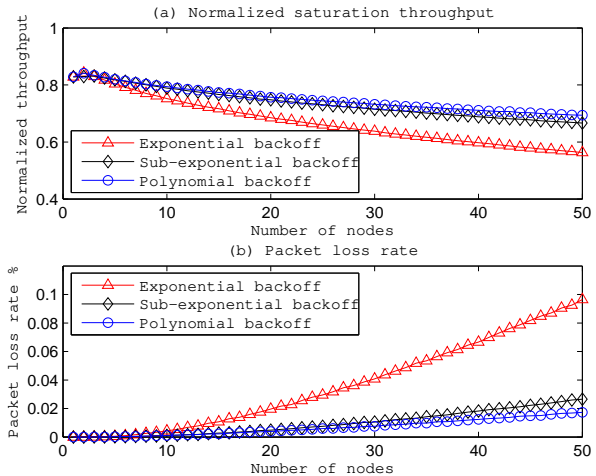}
  \end{center}
  \caption{Performance comparisons with retry limit $K=5$.}
  \label{56}
\end{figure}

Fig. $\ref{55}$ compares saturation throughput and delay performance of EB, SEB and PB. Here, the total simulation time is $10^6$ time slots. The backoff functions in consideration are EB ($g(k)=2^k$), SEB ($g(k)=4^{k^{0.7}}$) and PB ($g(k)=1+k^3$). The coefficients in the backoff functions are chosen to align the throughputs of different backoff schemes for fair comparison. No retry limit is imposed. When the network size increases from $1$ to $50$, normalized saturation throughput and the variance of medium access delay are plotted in Fig. $\ref{55}.a$ and Fig. $\ref{55}.b$, respectively. We can see that the three schemes yield similar saturation throughput. However, the delay variance of EB is much larger than those of SEB and PB. Specifically, the delay variance of EB is about $5$ times of PB. The root cause of this phenomenon is that the delay variance of EB is infinite following Theorem $1$ in Section IV (here $P_c> \frac{1}{r^2}$ when $N>8$). The delay variance of EB in Fig. $\ref{55}.b$ are bounded only due to finite simulation time. Similarly, we compare the three schemes in Fig. $\ref{56}$ when the retry limit $K=5$. The figure shows that EB perceives much higher packet loss rate than SEB and PB due to its inherent characteristics of power-law delay distribution. The higher packet loss rate of EB also causes notably lower throughput compared with SEB and PB. An intuitive explanation is that the delay distribution of EB has a ``heavier" tail than PB and SEB when $K=\infty$. By setting $K=5$, it is analogously truncating the delay distribution at a certain point and calculating the tail distribution beyond as the packet loss rate. Accordingly, EB with a ``heavier" tail also yields higher packet loss rate as well. Results in Fig. $\ref{55}$ and $\ref{56}$ suggest that SEB and PB achieve better delay performance than EB under the same network throughput, which is consistent with our results in Fig. $\ref{54}$. In particular, PB yields the smallest delay variance and lowest packet loss rate.

\begin{figure}
\centering
  \begin{center}
    \includegraphics[width=3.5in]{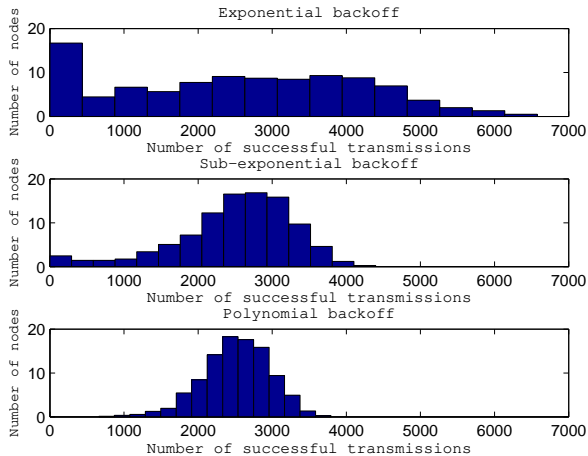}
  \end{center}
  \caption{Histogram of the number of successful transmissions under saturation condition.}
  \label{57}
\end{figure}

In Fig. $\ref{57}$, we compare the fairness among users when the three backoff schemes are used. The histogram of the number of successfully transmitted packets is plotted for $100$ nodes. The result is an average of $20$ independent simulations each with $10^7$ time slots. The average number of packets transmitted per node is $[2530,2491,2513]$ for EB, SEB and PB, respectively. This means that the throughput of the three backoff schemes are very similar. We can see in Fig. $\ref{57}.a$ that with EB, the difference in the amount of service received is significant across different nodes. The number of successfully transmitted packets by a node can vary all the way from $0$ to $6500$. Worse still, severe transmission starvation is observed with EB. On average more than $17\%$ of nodes transmit very few or even zero packets during the entire simulation time. In Fig. $\ref{57}.b$, SEB performs much better than EB, where the disparity of successful transmission is smaller and the maximum number of successfully transmitted packets is reduced to around $4000$. However, we can still observe around $3\%$ of the nodes in transmission starvation. This is because large delay can still occur with non-negligible probability when SEB is adopted, although the power law tail is mitigated. In vivid contrast, we can see in Fig. $\ref{57}.c$ that the range of the number of successful transmissions is significantly reduced and no transmission starvation occurs when PB is implemented. This indicates that PB achieves the fairest air time allocation among nodes. Such observation is consistent with our results in Fig. $\ref{54}$ that PB has the ``lightest" tail of delay distribution among the three schemes.

\subsection{Throughput performance of PB}
\begin{figure}
\centering
  \begin{center}
    \includegraphics[width=3.5in]{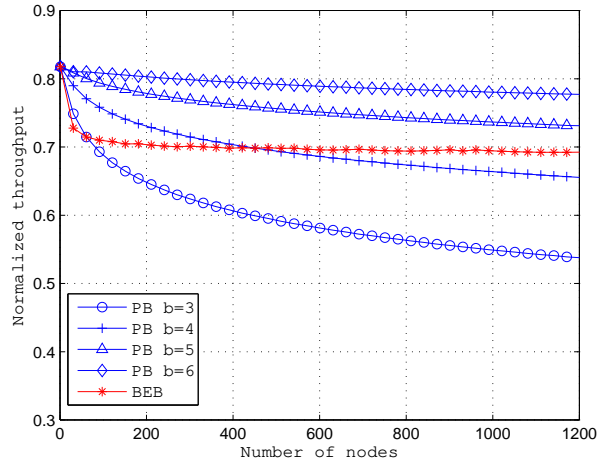}
  \end{center}
  \caption{Saturation throughput of PB ($g(k)=1+k^b$) when $b$ varies.}
  \label{58}
\end{figure}

With different backoff parameters, we plot the normalized saturation throughput of PB in Fig. $\ref{58}$ when the number of contending nodes varies from $1$ to $1200$. Besides, the saturation throughput of BEB is also presented for comparison. We can see that the throughput of PB gradually decreases as $N$ increases. In fact, the throughput will decrease to zero when $N$ becomes significantly large. On the other hand, the throughput of exponential backoff converges to a constant as $N$ increases. These observations verify out analysis in Theorem $2$ that the throughput is stable with EB while unstable with PB. However, we also see that PB with $b\geq5$ can sustain higher saturation throughput than BEB for all $ N \leq 1200$. This implies that high efficiency can be obtained with PB in practical scenarios when the order of backoff exponent is set properly. Therefore, we can safely enjoy the small delay jitter and better user fairness brought by PB without worrying about the instability of asymptotic throughput in practical systems.

Simulation results in this section show that PB can achieve high throughput, smaller delay jitter and good user fairness at the same time within practical range of user population. It is therefore a better alternative than EB, especially for carrying real-time traffics with stringent delay requirements.

\section{Conclusions}
In this paper, we have analyzed the tail delay distribution and throughput stability of general backoff functions. A tradeoff has been established between the tail decaying rate of medium access delay distribution and the stability of throughput. In particular, we found that power law delay distribution can be avoided if the backoff functions are ``slower" than an exponential function. Examples of such ``slow" backoffs are PB and SEB. In addition, the delay distribution becomes light tailed when linear-sublinear PB is used. On the other hand, non-zero asymptotic throughput is attainable only when backoff functions grow at least as fast as an exponential function, such as EB. For practical implementation, we show that PB obtains good throughput performance within a practical range of user population. Meanwhile, all delay moments with PB are finite as opposed to the infinite delay moments with EB. As such, we advocate PB as a better alternative than EB, now that there are increasingly more multimedia applications with stringent delay requirements in the network.

\appendices
\section{Proof of Corollary $1$}
\emph{Proof:} We first show that the limit
\begin{equation}
\label{70}
\lim_{j \rightarrow \infty}\frac{W_{j+1}^{n}}{\sum_{k=0}^{j}W_k^n} = \frac{1}{\sum_{k=0}^{j}\left[\frac{g(k)}{g(j+1)}\right]^n}
\end{equation}
exists. Since $\lim_{j\rightarrow \infty} \frac{g(j+1)}{g(j)} = \gamma$, it holds that $\forall \epsilon>0$, $\exists k_0>0$ such that
\begin{equation}
\left|\frac{g(k+1)}{g(k)}-\gamma\right|< \epsilon,\ \ \forall k>k_0.
\end{equation}
Equivalently, we have
\begin{equation}
\frac{1}{\gamma + \epsilon}<\frac{g(k)}{g(k+1)}<\frac{1}{\gamma-\epsilon},\ \ \forall k>k_0.
\end{equation}
Then, it holds that
\begin{equation}
\left(\frac{1}{\gamma+\epsilon}\right)^{j-k}<\frac{g(k)}{g(j)}<\left(\frac{1}{\gamma-\epsilon}\right)^{j-k},\ \ \forall j>k>k_0.
\end{equation}

The second term of the RHS of (\ref{70}) can be rewritten as
\begin{equation}
\label{71}
\frac{1}{\sum_{k=0}^{j}\left[\frac{g(k)}{g(j+1)}\right]^n} = \frac{1}{\sum_{k=0}^{k_0}\left[\frac{g(k)}{g(j+1)}\right]^n + \sum_{k=k_0+1}^{j}\left[\frac{g(k)}{g(j+1)}\right]^n},
\end{equation}
which is upper bounded by
\begin{equation}
\begin{aligned}
&\frac{1}{\sum_{k=0}^{k_0}\left[\frac{g(k)}{g(j+1)}\right]^n + \sum_{k=1}^{j-k_0}\left(\frac{1}{\gamma+\epsilon}\right)^{kn}}
\end{aligned}
\end{equation}
and lower bounded by
\begin{equation}
\label{76}
\begin{aligned}
&\frac{1}{\sum_{k=0}^{k_0}\left[\frac{g(k)}{g(j+1)}\right]^n + \sum_{k=1}^{j-k_0}\left(\frac{1}{\gamma-\epsilon}\right)^{kn}}.
\end{aligned}
\end{equation}
Notice that $\epsilon$ ($g(j+1)$) can be made arbitrarily small (large) as $j\rightarrow \infty$. When $\gamma>1$, we see that the lower bound and upper bounds converge to $\gamma^n-1$ as $j\rightarrow \infty$. Otherwise, when $\gamma=1$, the lower bound and upper bounds converge to $0$ as $j\rightarrow \infty$. In both cases,
\begin{equation}
\label{63}
\lim_{j\rightarrow\infty}\frac{W_{j+1}^{n}}{\sum_{k=0}^{j}W_k^n} = \gamma^n-1.
\end{equation}

Next, we prove that $E\left[\Lambda^n\right]$ is finite only if $ P_c\gamma^n<1$. From (\ref{22}), we have
\begin{equation}
\label{75}
\begin{aligned}
E[\Lambda^n]\geq\frac{1}{2^n}\left(1-P_c\right)\sum_{j=0}^{\infty}\left\{P_c^j \sum_{k=0}^{j}\left(W_k-1\right)^n\right\}.
\end{aligned}
\end{equation}
Letting
\begin{equation}
U_j = P_c^j \sum_{k=0}^{j}\left(W_k-1\right)^n,
\end{equation}
we see that the limit
\begin{equation}
\lim_{j\rightarrow \infty}\frac{U_{j+1}}{U_j} = P_c \left(1+\frac{\left(W_{j+1}-1\right)^n}{\sum_{k=0}^{j}\left(W_k-1\right)^n}\right)
\end{equation}
exists following the similar argument in (\ref{70})-(\ref{76}). Using the ratio test, the lower bound in (\ref{75}) is finite only if
\begin{equation}
\label{41}
\begin{aligned}
\lim_{j\rightarrow \infty} \frac{U_{j+1}}{U_j} = P_c \left(1+\frac{\left(W_{j+1}-1\right)^n}{\sum_{k=0}^{j}\left(W_k-1\right)^n}\right)\leq 1.
\end{aligned}
\end{equation}
Since $W_{k+1}>W_{k}>1$, we have
\begin{equation}
\frac{\sum_{k=0}^{j}\left(W_k-1\right)^n}{\left(W_{j+1}-1\right)^n} < \frac{\sum_{k=0}^{j}\left(W_k\right)^n}{\left(W_{j+1}\right)^n},\ \forall j.
\end{equation}
Equivalently, it holds that
\begin{equation}
\label{42}
\frac{\left(W_{j+1}-1\right)^n}{\sum_{k=0}^{j}\left(W_k-1\right)^n} > \frac{\left(W_{j+1}\right)^n}{\sum_{k=0}^{j}\left(W_k\right)^n},\ \forall j.
\end{equation}
With (\ref{41}) and (\ref{42}), the lower bound converges only if
\begin{equation}
\begin{aligned}
\lim_{j\rightarrow \infty} P_c \left(1+\frac{W_{j+1}^n}{\sum_{k=0}^{j}W_k^n}\right)< 1.
\end{aligned}
\end{equation}
From (\ref{63}), the above condition can be rewritten as $P_c\gamma^n<1$. We therefore reach the proof of \emph{only if} part, since a finite lower bound is a necessary condition for finite $E[\Lambda^n]$.

Then, we prove that $E\left[\Lambda^n\right]$ is finite if $P_c\gamma<1$. From (\ref{74}), we have
\begin{equation}
E[\Lambda^n]\leq\frac{1-P_c}{n+1}\sum_{j=0}^{\infty}\left\{P_c^j\left(j+1\right)^{n-1}\cdot \sum_{k=0}^{j}W_k^{n}\right\}.
\end{equation}
Similarly to the argument in the proof of the \emph{only if} argument, the following limit exists, where
\begin{equation}
\begin{aligned}
&\lim_{j\rightarrow \infty} \frac{P_c^{j+1}\left(j+2\right)^{n-1}\cdot\sum_{k=0}^{j+1}W_k^n}{P_c^j\left(j+1\right)^{n-1}\cdot\sum_{k=0}^{j}W_k^n}\\
=&\lim_{j\rightarrow \infty} P_c\cdot \left(1+\frac{W_{j+1}^{n}}{\sum_{k=0}^{j}W_k^n}\right) = P_c\gamma^n.
\end{aligned}
\end{equation}
Using the ratio test, the upper bound in (\ref{74}) converges if $P_c\gamma^n<1$. A convergent upper bound is a sufficient condition for finite $E[\Lambda^n]$, which leads to the proof that $E\left[\Lambda^n\right]$ is finite.

Notice that $P_c<1$ strictly holds in a random access network with finite number of nodes. Therefore, $P_c \gamma^n <1$ holds for all $n\in \mathbb{N}$ when $\gamma=1$. When $\gamma>1$, however, $P_c \gamma^n <1$ is violated for all $n \geq -\frac{\ln P_c}{\ln \gamma}$. In other words, there always exists infinite delay moment when $\gamma>1$, while all the delay moments are finite when $\gamma=1$. Following the definition of power law distribution, we reach the proof of Corollary $1$. $\hfill \blacksquare$

\section{Heavy-tailed Behavior of Delay Distribution}
As illustrated in Fig. $\ref{53}$, power law distribution belongs to a subclass of heavy-tailed distribution (region $2$ in Fig. $\ref{53}$). However, the converse statement, that a non-power law distribution is not heavy-tailed, is not true (region $3$). This motivates us to further study the heavy-tailed behavior of medium access delay distribution of different backoff functions. Interestingly, we show that superlinear PB and SEB, which eliminate the power law tail, still yield heavy-tailed delay distribution (region $3$). Meanwhile, linear-sublinear PB, i.e., $g(k)=1+k^b$ and $0<b\leq1$, yields a light-tailed delay distribution (region $4$).

\subsection{Polynomial backoff}
Since medium access delay distribution is a power law distribution with EB, it is also a heavy-tailed distribution. For PB, we show that the delay distribution is heavy-tailed if $b>1$, whereas lighted-tailed if $0<b\leq1$.

From (\ref{16}), it holds that
\begin{equation}
\text{E}\left[X^n\right] \geq \text{E}\left[\Lambda^n\right]\cdot L_{\text{min}}^n,
\end{equation}
where $L_{\text{min}}=\min\{T_{idle},T_{succ},T_{coll}\}$. From (\ref{22}), we have
\begin{equation}
\label{45}
\begin{aligned}
\text{E}\left[\Lambda^n\right]\geq \left(1-P_c\right)\sum_{j=0}^{\infty}\left\{P_c^j \sum_{k=0}^{j}\left(\text{E}\left[B_k\right]\right)^n\right\}.
\end{aligned}
\end{equation}
For PB, we substitute $E\left[B_k\right]=\frac{1}{2}\left(1+k^b\right)W_0$ into the RHS of (\ref{45}),
\begin{equation}
\begin{aligned}
\text{E}\left[\Lambda^n\right] &\geq \left(1-P_c\right)\left(\frac{W_0}{2}\right)^n\sum_{j=0}^{\infty}\left\{P_c^j \sum_{k=0}^{j}\left(1+k^b\right)^n\right\}\\
&\geq \left(1-P_c\right)\left(\frac{W_0}{2}\right)^n\sum_{j=0}^{\infty}\left\{P_c^j \sum_{k=0}^{j}k^{bn}\right\}\\
&\geq \left(1-P_c\right)\left(\frac{W_0}{2}\right)^n\sum_{j=0}^{\infty}\left\{P_c^j \int_{0}^{j} t^{bn} dt\right\}\\
&= \frac{1-P_c}{bn+1}\left(\frac{W_0}{2}\right)^n\sum_{j=0}^{\infty}P_c^j j^{bn+1}.
\end{aligned}
\end{equation}
Here, $\sum_{j=0}^{\infty}P_c^j j^{bn+1}$ can be represented as
\begin{equation}
\label{47}
\sum_{j=0}^{\infty}P_c^j j^{bn+1} \triangleq \Phi \left(P_c, -(bn+1), 0\right),
\end{equation}
where
\begin{equation}
\Phi \left(z, s, v\right) = \sum_{i=0}^{\infty}z^{i}\left(v+i\right)^{-s}
\end{equation}
is a Lerch's transcendent. When $s<0$ and $|z|<1$, it holds that
\begin{equation}
\begin{aligned}
\Phi \left(z, s, v\right) \approx \frac{\Gamma{(1-s)}}{z^v}\left(\ln\frac{1}{z}\right)^{s-1}
\end{aligned}
\end{equation}
(cf. \cite{1953:Bateman}, p. $29$), where $\Gamma(x)$ is a Gamma function. Correspondingly, we can write
\begin{equation}
\label{31}
\begin{aligned}
\text{E}\left[\Lambda^n\right]&\geq \frac{1-P_c}{bn+1}\left(\frac{W_0}{2}\right)^n\Phi \left(P_c, -(bn+1), 0\right)\\
&\approx\frac{1-P_c}{bn+1}\left(\frac{W_0}{2}\right)^n\Gamma(bn+2) \left(\ln\frac{1}{P_c}\right)^{-\left(bn+2\right)}.
\end{aligned}
\end{equation}

Recall in (\ref{26}) that $f(x)$ is heavy-tailed if and only if
\begin{equation}
\label{38}
\sum_{n=0}^{\infty} \frac{\lambda^n}{n!}\text{E}\left[X^n\right]=\infty,\ \ \forall \lambda>0.
\end{equation}
We substitute the inequality in (\ref{31}) into the LHS of (\ref{38}),
\begin{equation}
\label{48}
\begin{aligned}
&\sum_{n=0}^{\infty} \frac{\lambda^n}{n!}\text{E}\left[X^n\right] \\
\geq&  \sum_{n=0}^{\infty} \frac{\lambda^n}{n!} \text{E}\left[\Lambda^n\right]L_{\text{min}}^n \\
\geq&  \sum_{n=0}^{\infty} \frac{1-P_c}{bn+1}\left(\frac{W_0L_{\text{min}}\lambda}{2}\right)^n\frac{1}{n!} \Gamma(bn+2) \left(\ln\frac{1}{P_c}\right)^{-\left(bn+2\right)}.
\end{aligned}
\end{equation}
The LHS of (\ref{38}) is finite only if the lower bound is finite. Using the ratio test to lower bound in (\ref{48}), we obtain the test parameter as
\begin{equation}
\label{30}
\begin{aligned}
\Delta(\lambda) = \frac{W_0L_{\text{min}}\lambda}{2}\ln\left(\frac{1}{P_c}\right)^{-b} \cdot \lim_{n\rightarrow \infty} \frac{\Gamma(bn+2+b)}{\Gamma(bn+2)(n+1)}.
\end{aligned}
\end{equation}
The delay distribution of PB is heavy-tailed if $\Delta(\lambda)$ is larger than $1$ for all $\lambda>0$.

Note that, when $x$ is a very large real positive number, Gamma function can be well approximated by
\begin{equation}
\ln\Gamma(x) \approx (x-1/2) \ln x - x + \frac{1}{2}\ln(2\pi)
\end{equation}
(cf. \cite{1953:Bateman}, p. $21$). We denote $y\triangleq bn+2$, then
\begin{equation}
\label{69}
\begin{aligned}
&\lim_{n\rightarrow \infty}\ln\left(\frac{\Gamma(bn+2+b)}{\Gamma(bn+2)(n+1)}\right)\\
=& \lim_{y,n\rightarrow \infty}\left\{\ln\Gamma(y+b) - \ln\Gamma(y)- \ln(n+1)\right\}\\
\approx& \lim_{y,n\rightarrow \infty}\left\{\left(y+b-\frac{1}{2}\right)\ln(y+b) - (y+b) \right.\\
&\left.- \left(y-\frac{1}{2}\right)\ln y + y -\ln(n+1)\right\}\\
=& \lim_{y,n\rightarrow \infty}\biggr\{\left(y-\frac{1}{2}\right)\ln\left(\frac{y+b}{y}\right) + b\ln (y+b) - \ln(n+1) -b\biggr\}\\
=& \lim_{y,n\rightarrow \infty}\left\{y\ln\left(\frac{y+b}{y}\right) + b\ln (y+b) - \ln(n+1) -b\right\}.\\
\end{aligned}
\end{equation}
By L'Hospital's rule, $\lim_{y\rightarrow \infty}y\ln\left(\frac{y+b}{y}\right)=b$. Then, the RHS of (\ref{69}) equals to
\begin{equation}
\begin{aligned}
&\lim_{y,n\rightarrow \infty}\left\{b\ln (y+b) - \ln(n+1)\right\}\\
=& \lim_{n\rightarrow \infty} \ln\left(\frac{(bn+2+b)^b}{n+1}\right)\\
=& \lim_{n\rightarrow \infty} \ln\left(\frac{(bn)^b}{n}\right)= \lim_{n\rightarrow \infty} \left\{ (b-1)\ln n +b\ln b\right\}.
\end{aligned}
\end{equation}
Therefore, we have
\begin{equation}
\label{28}
\lim_{n\rightarrow \infty}\frac{\Gamma(bn+2+b)}{\Gamma(bn+2)(n+1)}=
\begin{cases}
0, &b<1,\\
1,&b=1,\\
\infty, &b>1.\\
\end{cases}
\end{equation}
Accordingly, the test parameter $\Delta(\lambda)$ in (\ref{30}) is
\begin{equation}
\Delta(\lambda)=
\begin{cases}
0, &b<1,\\
\frac{W_0L_m\lambda}{2}\ln\left(\frac{1}{P_c}\right)^{-b}, &b=1,\\
\infty, &b>1.\\
\end{cases}
\end{equation}

We can see that $\Delta(\lambda)=\infty$ for all $\lambda$ when $b>1$. Therefore, the delay distribution of a super-linear PB is heavy-tailed. For linear-sublinear PB with $b\leq 1$, however, we currently have not obtained conclusive analytical results to verify its heavy-tailed behavior.

\begin{figure}
\centering
  \begin{center}
    \includegraphics[width=3.5in]{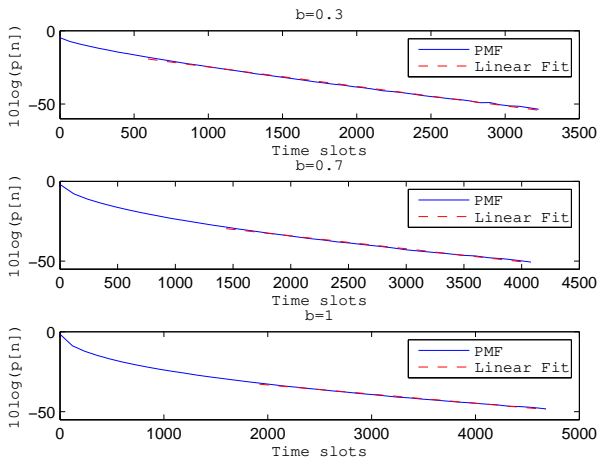}
  \end{center}
  \caption{Probability mass function (PMF) of $\Lambda$ with different $b$. The $R^2$ values of linear regressions are $0.9976$, $0.9956$ and $0.9951$ for the three cases, respectively.}
  \label{59}
\end{figure}

Instead, we numerically calculate the probability mass function of $\Lambda$ and find it matches the features of light-tailed distribution when $0<b\leq1$. The probability mass function can be obtained through calculating its probability generating function (cf. \cite{2006:Mieghem}, p. $33$). We plot $\log(p[n])$ against $n$ with different $0<b\leq1$ in Fig. $\ref{59}$. Besides, the results of linear regressions and corresponding $R^2$ values are also provided. In all three cases, we can see that $\log(p[n])\thicksim -\lambda_0 n$ for some $\lambda_0>0$, indicating $p[n]\thicksim e^{-\lambda_0 n}$. Thus, the delay distribution is light-tailed distribution with an exponentially decaying tail.

\subsection{Sub-exponential backoff}
It can be easily shown that SEB also suffers heavy-tail delay distribution. To see this, we substitute $\text{E}\left[B_k\right]=\frac{1}{2}r^{k^a}W_0$ into (\ref{45}), then
\begin{equation}
\label{34}
\begin{aligned}
\text{E}\left[\Lambda^n\right]\geq \left(1-P_c\right)\left(\frac{W_0}{2}\right)^n\sum_{j=0}^{\infty}\left\{P_c^j \sum_{k=0}^{j}r^{nk^a}\right\}.
\end{aligned}
\end{equation}
For any $0<a<1$, $r>1$ and $b>1$, there always $\exists m\in\mathbb{N}$ so that
\begin{equation}
\sum_{k=0}^{j}r^{nk^a}>\sum_{k=0}^{j}\left(1+k^b\right)^n,\ \ \forall j\geq m.
\end{equation}
As we have proved in previous subsection, the following inequality holds for PB with $f(k)=1+k^b$ and $b>1$,
\begin{equation}
\label{35}
\sum_{n=0}^{\infty}\frac{\lambda^n}{n!}\left[\left(1-P_c\right)\left(\frac{W_0}{2}\right)^n\sum_{j=0}^{\infty}\left\{P_c^j \sum_{k=0}^{j}\left(1+k^b\right)^n\right\}\right]=\infty,
\end{equation}
$\forall \lambda>0$. Following (\ref{34}) and (\ref{35}), it suffices to claim to that the moment generating function of SEB diverges, i.e.,
\begin{equation}
\int_0^{\infty} e^{\lambda x}f(x)dx=\sum_{n=0}^{\infty}\frac{\lambda^n}{n!}\text{E}\left[X^n\right]=\infty,\ \  \forall \lambda>0.
\end{equation}
Therefore, the distribution of medium access delay is heavy-tailed.

\end{document}